# A Bi-Layer TSN Formulation for Separable Scheduling of Mobile Emergency Resources

Wei Wang, *Member, IEEE*, Minwu Chen, *Senior Member, IEEE*, Yinyu Chen, *Member, IEEE*, Jingyan He

***Abstract*—Separable scheduling unleashes the deployment flexibility of mobile emergency resources by dispatching carriers and functional modules separately yet in a coordinated manner, offering a promising avenue to enhance power system resilience. However, this flexibility induces a distinct carrier-supported module routing structure, where non-self-mobile modules must be routed through compatible carrier movements. The resulting carrier-module spatio-temporal coupling makes exact and tractable optimization challenging. This letter identifies this structure and develops a novel exact bi-layer time-space network formulation as a mixed-integer linear program. The proposed formulation represents carrier and module trajectories as interacting network flows and enforces their support relations through explicit arc-level coupling. Compared with the prior logic-based model, the proposed formulation preserves exactness while improving modeling flexibility by eliminating mandatory post-arrival dwelling. Numerical studies validate its correctness and demonstrate substantial computational advantages.**

## I. Introduction

AMID the rising threat of extreme weather events, mobile emergency resources (MERs), such as truck-, trailer-, rail-mounted, and other transportable energy storage systems and generators, have become pivotal for enhancing power grid resilience [1], [2]. By moving supply and recovery capabilities to affected areas, MERs provide spatial flexibility beyond fixed infrastructure. Harnessing this flexibility, however, requires tractable spatio-temporal scheduling models to coordinate MER movements and service decisions, which remains challenging in resilience-oriented power system operation.

Existing MER scheduling models predominantly adopt logic-based or time-space network (TSN) formulations. Logic-based formulations characterize MER movements and service decisions through time-indexed logical relations and transition constraints [3], [4], [5], whereas TSN formulations represent the spatio-temporal evolution of MERs as arc flows on a time-expanded network [6], [7]. Recent studies further highlight the strength of TSN formulations in coordinating heterogeneous mobile units [8]. Nevertheless, existing models typically treat MERs as integrated self-mobile entities, leaving the potential arising from carrier-module separability largely untapped.

Breaking the rigid bundling of carriers and functional modules (*e.g.*, carried generators and energy storage units) unlocks a new dimension of flexibility for MER deployment. This motivates the concept of separable scheduling, introduced in our previous work [9], which leverages the physical separability of many commercial MERs (*e.g.*, tractor-trailer configurations) to dispatch carriers and modules separately yet in a coordinated manner. The resulting expanded scheduling flexibility, however, gives rise to more complex carrier-module spatio-temporal coupling: carriers and modules may follow distinct trajectories, while any module displacement must be supported by a compatible carrier movement. As a result, exact optimization requires simultaneously tracking carrier trajectories, module trajectories, and their support relations. The previous logic-based formulation in [9] captures this intricate coupling, but its high computational burden limits its applicability in time-critical resilience-oriented scheduling. Moreover, its mandatory post-arrival dwelling requirement may restrict operational flexibility. A new formulation is therefore called for to improve computational tractability while retaining an exact representation of separable scheduling.

To this end, this letter develops a novel exact bi-layer TSN formulation for separable MER scheduling. By recasting it as a carrier-supported module routing problem, the proposed formulation casts carrier and module movements as interacting flows on interdependent TSN layers and captures their spatio-temporal coupling through explicit arc-support links. The resulting formulation retains exactness while substantially boosting computational efficiency, thereby advancing separable scheduling toward a more scalable and practical framework for fully exploiting MER flexibility.

## II. The Bi-Layer TSN Formulation

In separable scheduling, carriers and modules are separately routed and recombined for service delivery at demand locations, as shown in Fig. 1. This expands the MER scheduling decision space, with conventional integrated dispatch subsumed as a restricted special case [9]. Since non-self-mobile modules can move only through carrier support, separable scheduling is naturally formalized as a carrier-supported module routing problem over interdependent TSN layers.

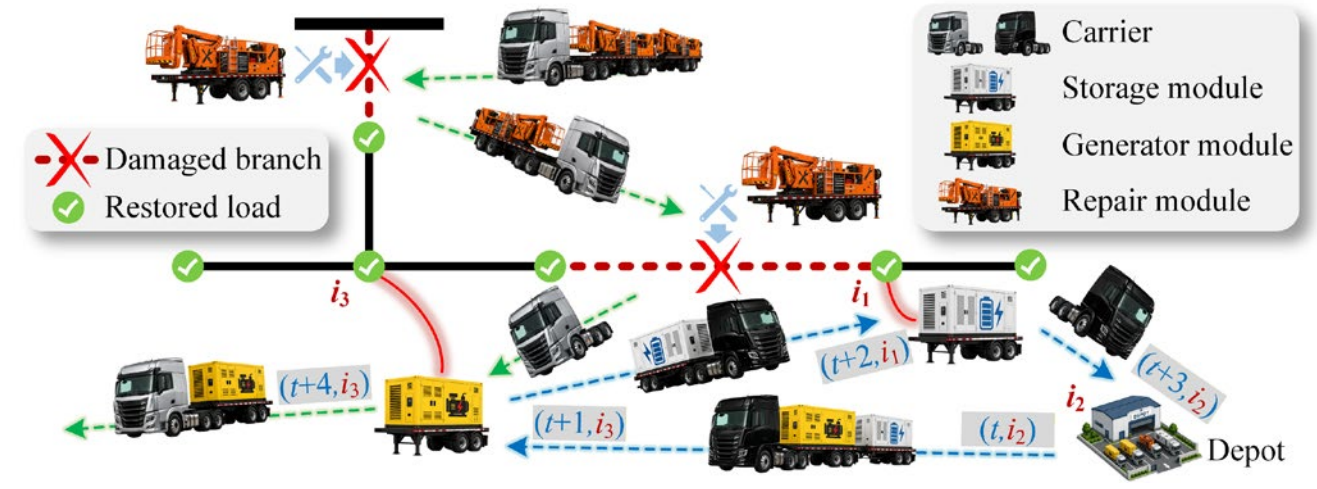

Fig. 1. Separable MER scheduling with multiple module types.

### A. Bi-layer TSN Structure and Notation

Accordingly, a bi-layer TSN structure is built with two interdependent classes of TSNs: carrier- and module-side TSNs, as shown in Fig. 2. Let $\mathcal{J}$ and $\mathcal{K}$ denote the sets of carriers and modules, respectively. Each carrier $j\in\mathcal{J}$ has a carrier-side TSN with hold and travel arcs for staying and traveling, respectively. Similarly, each module $k\in\mathcal{K}$ has a module-side TSN with hold arcs for staying and transport arcs for carrier-supported movements. This induces an arc-support coupling between the two classes of TSNs: a module transport arc can be activated only if the compatible carrier travel arc is selected.

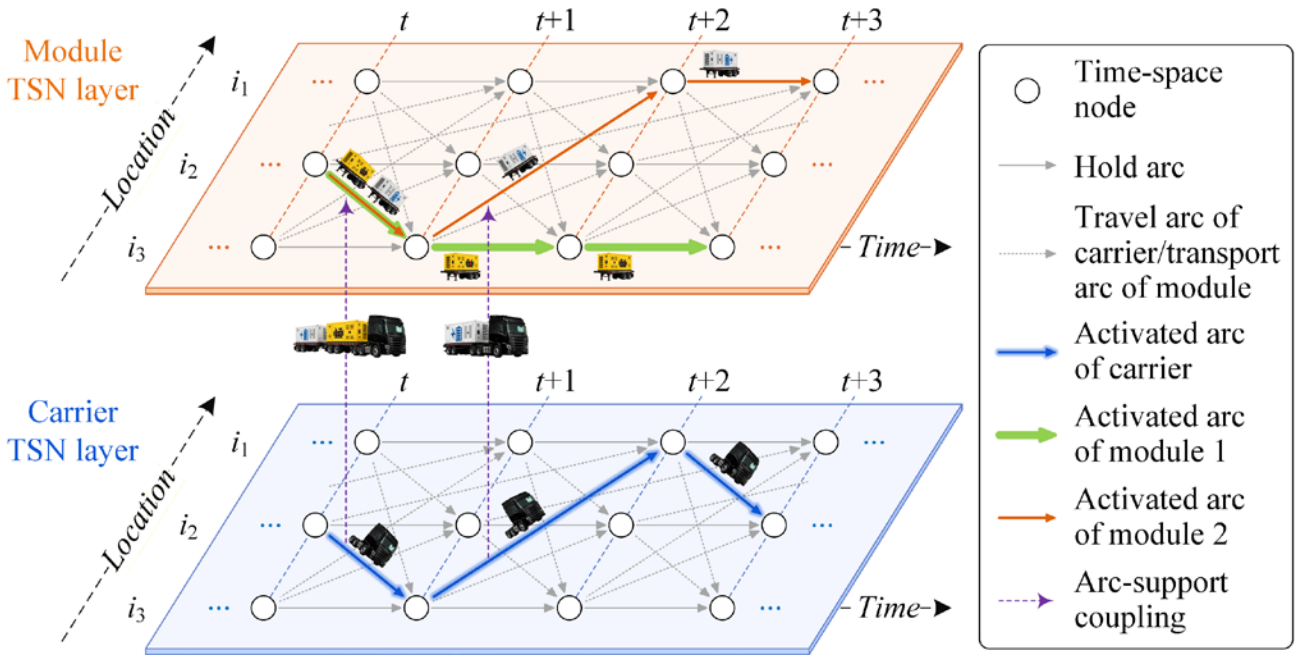

Fig. 2. Illustration of the bi-layer TSN for separable MER scheduling.

Let $\mathcal{N}$ be the set of physical locations and $\mathcal{T}=\{0, 1, 2, \ldots, D\}$ be the set of time points. A TSN node is defined as a time-location pair $(t, i)$, where $t\in\mathcal{T}$ and $i\in\mathcal{N}$. Therefore, the carrier- and module-side TSNs are constructed using the four types of arcs summarized in Table I.

TABLE I. Definitions of Carrier- and Module-Side TSN Arc Sets

| Object | Description | Definition |
|---|---|---|
| Carrier $j\in\mathcal{J}$ | *Hold arcs*: stays at $i$ | $\mathcal{A}_j^{\mathrm{H}}=\{((t,i),(t+1,i)): t\in\mathcal{T}\backslash\{D\}, i\in\mathcal{N}_j^{\mathrm{C}}\}$ |
| | *Travel arcs*: moves from $i$ to $i'$ | $\mathcal{A}_j^{\mathrm{T}}=\{((t,i),(t+\tau_{j,ii'},i')): t\in\mathcal{T}, t+\tau_{j,ii'}\le D, i,i'\in\mathcal{N}_j^{\mathrm{C}}, i\neq i'\}$ |
| Module $k\in\mathcal{K}$ | *Hold arcs*: stays at $i$ | $\mathcal{B}_k^{\mathrm{H}}=\{((t,i),(t+1,i)): t\in\mathcal{T}\backslash\{D\}, i\in\mathcal{N}_k^{\mathrm{M}}\}$ |
| | *Transport arcs*: transported from $i$ to $i'$ by carrier $j$ | $\mathcal{B}_{k,j}^{\mathrm{T}}=\{((t,i),(t+\tau_{j,ii'},i')): ((t,i),(t+\tau_{j,ii'},i'))\in\mathcal{A}_j^{\mathrm{T}}, i,i'\in\mathcal{N}_k^{\mathrm{M}}, \rho_{k,j}=1\}$ |

where $\mathcal{N}_j^{\mathrm{C}}\subseteq\mathcal{N}$ and $\mathcal{N}_k^{\mathrm{M}}\subseteq\mathcal{N}$ denote the feasible location sets of carrier $j$ and module $k$, respectively, which may coincide in most cases. $\tau_{j,ii'}$ is the travel time of carrier $j$ from location $i$ to $i'$, and $\rho_{k,j}=1$ indicates that carrier $j$ can transport module $k$. Therefore, feasible carrier and module trajectories are encoded as paths over their corresponding arcs in the TSNs.

## B. Carrier-Supported Module Routing Formulation

Based on the bi-layer TSN, separable MER scheduling is formulated as the following carrier-supported module routing MILP within a network-flow framework. The main notation, in addition to Table I, is summarized in Table II.

TABLE II. Main Notation

| | | | |
|---|---|---|---|
| ***Sets*** | | | |
| $\delta_j^-$ / $\delta_j^+$ | In/out arcs of carrier $j$ at a TSN node | $\delta_j^{\mathrm{T},-}$ | Incoming travel arcs of carrier $j$ |
| $\delta_k^{\mathrm{H},-}$ / $\delta_k^{\mathrm{H},+}$ | In/out hold arcs of module $k$ | $\delta_{k,j}^{\mathrm{T},-}$ / $\delta_{k,j}^{\mathrm{T},+}$ | In/out transport arcs of module $k$ on carrier $j$ |
| ***Variables*** | | | |
| $y_{j,a}$ | Binary. 1 if arc $a$ of carrier $j$ is activated, $a\in\mathcal{A}_j=\mathcal{A}_j^{\mathrm{H}}\cup\mathcal{A}_j^{\mathrm{T}}$ | $x_{k,a}$ | Binary. 1 if hold arc $a$ of module $k$ is activated, $a\in\mathcal{B}_k^{\mathrm{H}}$ |
| $z_{k,j,a}$ | Binary. 1 if transport arc $a$ of module $k$ on carrier $j$ is activated, $a\in\mathcal{B}_{k,j}^{\mathrm{T}}$ | $s_{i,t}$ | Served demand at location $i$ during $(t-1, t)$ |
| ***Parameters*** | | | |
| $Q_j$/ $q_k$ | Transport capacity of carrier $j$/ size of module $k$ | $p_k$ | Service capacity of module $k$ |
| $d_{i,t}$ | Demand at location $i$ during $(t-1,t)$ | $w_i$ | Service priority weight of location $i$ |
| $c_j$ | Unit travel cost of carrier $j$ | $\lambda$ | Penalty weight for travel |

The routing constraints are first established for carriers. Each carrier starts from its initial location, and subsequent trajectory is enforced by flow conservation over the carrier-side TSN:

$$\sum_{a\in\delta_j^+(0,i_j^0)} y_{j,a}=1 \;,\; \forall j\in\mathcal{J} \tag{1}$$

$$\sum_{i\in\mathcal{N}_j^{\mathrm{C}}\backslash\{i_j^0\}}\sum_{a\in\delta_j^+(0,i)} y_{j,a}=0 \;,\; \forall j\in\mathcal{J} \tag{2}$$

$$\sum_{a\in\delta_j^-(t,i)} y_{j,a}=\sum_{a\in\delta_j^+(t,i)} y_{j,a} \;,\; \forall j\in\mathcal{J}, i\in\mathcal{N}_j^{\mathrm{C}}, t\in\mathcal{T}\backslash\{0,D\} \tag{3}$$

where $i_j^0\in\mathcal{N}_j^{\mathrm{C}}$ is the initial location of carrier $j$. Constraints (1)-(2) determine starting node, while (3) ensures path continuity.

Module routing is formulated similarly, but with a key difference: non-self-mobile modules can only stay through hold arcs or move through carrier-supported transport arcs

$$\sum_{a\in\delta_k^{\mathrm{H},+}(0,i_k^0)} x_{k,a}+\sum_{j\in\mathcal{J}}\sum_{a\in\delta_{k,j}^{\mathrm{T},+}(0,i_k^0)} z_{k,j,a}=1 \;,\; \forall k\in\mathcal{K} \tag{4}$$

$$\sum_{i\in\mathcal{N}_k^{\mathrm{M}}\backslash\{i_k^0\}}\left(\sum_{a\in\delta_k^{\mathrm{H},+}(0,i)} x_{k,a}+\sum_{j\in\mathcal{J}}\sum_{a\in\delta_{k,j}^{\mathrm{T},+}(0,i)} z_{k,j,a}\right)=0 \;,\; \forall k\in\mathcal{K} \tag{5}$$

$$\begin{gathered}\sum_{a\in\delta_k^{\mathrm{H},-}(t,i)} x_{k,a}+\sum_{j\in\mathcal{J}}\sum_{a\in\delta_{k,j}^{\mathrm{T},-}(t,i)} z_{k,j,a}= \\ \sum_{a\in\delta_k^{\mathrm{H},+}(t,i)} x_{k,a}+\sum_{j\in\mathcal{J}}\sum_{a\in\delta_{k,j}^{\mathrm{T},+}(t,i)} z_{k,j,a}\;, \\ \forall k\in\mathcal{K}, i\in\mathcal{N}_k^{\mathrm{M}}, t\in\mathcal{T}\backslash\{0,D\}\end{gathered} \tag{6}$$

where $i_k^0\in\mathcal{N}_k^{\mathrm{M}}$ is the initial location of module $k$. Constraints (4)-(6) balance hold and transport arcs to route each module without introducing any self-mobile path.

The above constraints construct carrier and module paths separately, while their physical coordination is enforced through arc-support coupling. Specifically, a module transport arc cannot be activated unless the corresponding carrier travel arc is selected, and the total size of modules transported on that arc cannot exceed the carrier capacity:

$$z_{k,j,a}\le y_{j,a} \;,\; \forall k\in\mathcal{K}, j\in\mathcal{J}, a\in\mathcal{B}_{k,j}^{\mathrm{T}} \tag{7}$$

$$\sum_{k:\,a\in\mathcal{B}_{k,j}^{\mathrm{T}}} q_k z_{k,j,a}\le Q_j y_{j,a} \;,\; \forall j\in\mathcal{J}, a\in\mathcal{A}_j^{\mathrm{T}} \tag{8}$$

Although (7) is redundant for integer feasibility given (8) and the binary domains, it is retained to strengthen the LP relaxation. Moreover, the proposed TSN formulation avoids the mandatory post-arrival dwelling imposed in [9].

## C. MER Scheduling Formulation

To form a complete scheduling model, we adopt a simplified service representation that links served demand to modules staying at demand locations. Detailed service delivery models for MERs are application-dependent and can be adapted from existing MER scheduling studies [3]-[9], while this letter focuses on the carrier-supported routing structure of separable MERs and the validation of the bi-layer TSN formulation.

Specifically, the served demand at each location and time interval is bounded by the local demand and the service capacity of modules staying there:

$$0 \le s_{i,t} \le d_{i,t} \quad , \ \forall i \in \mathcal{N}, t \in \mathcal{T} \setminus \{0\} \tag{9}$$

$$s_{i,t} \le \sum_{k \in \mathcal{K}} \sum_{a \in \mathcal{B}_k^{\mathrm{H}} : a = ((t-1,i),(t,i))} p_k x_{k,a} \quad , \ \forall i \in \mathcal{N}, t \in \mathcal{T} \setminus \{0\} \tag{10}$$

The objective maximizes the weighted served demand and penalizes carrier travel costs:

$$\max \sum_{i \in \mathcal{N}} \sum_{t \in \mathcal{T} \setminus \{0\}} w_i s_{i,t} - \lambda \sum_{j \in \mathcal{J}} c_j \sum_{a \in \mathcal{A}_j^T} \tau_{j,a} y_{j,a} \tag{11}$$

where $\tau_{j,a}$ is the travel time associated with arc $a$ of carrier $j$.

## III. Case Studies

Case studies are conducted to verify the correctness and computational efficiency of the proposed formulation, with the prior logic-based model in [9] used as the benchmark. All tests are implemented in AMPL and solved by Gurobi 13.0 on a workstation with an Intel Xeon w7-3455 CPU and 128 GB RAM. The time limit is set to 3600 s and the *mipgap* to $10^{-4}$.

Fig. 3 presents an 8-location, 12-period instance where all carriers and modules depart from location 4. The demand is constructed to shift from locations 1-3 in early periods to locations 5-6 and finally to 7-8. Each module provides 1 unit of service per period. Each carrier has a transport capacity of 2, and the sizes are set to 1, 1, 2, 1, 2, and 2 for modules 1-6. Travel times are set to 2 among locations 6-8 and 1 otherwise. The results show clear demand-following and reconfigurable deployment behaviors, confirming that the proposed bi-layer TSN faithfully represents carrier-supported routing of separable MERs under capacity and coupling constraints.

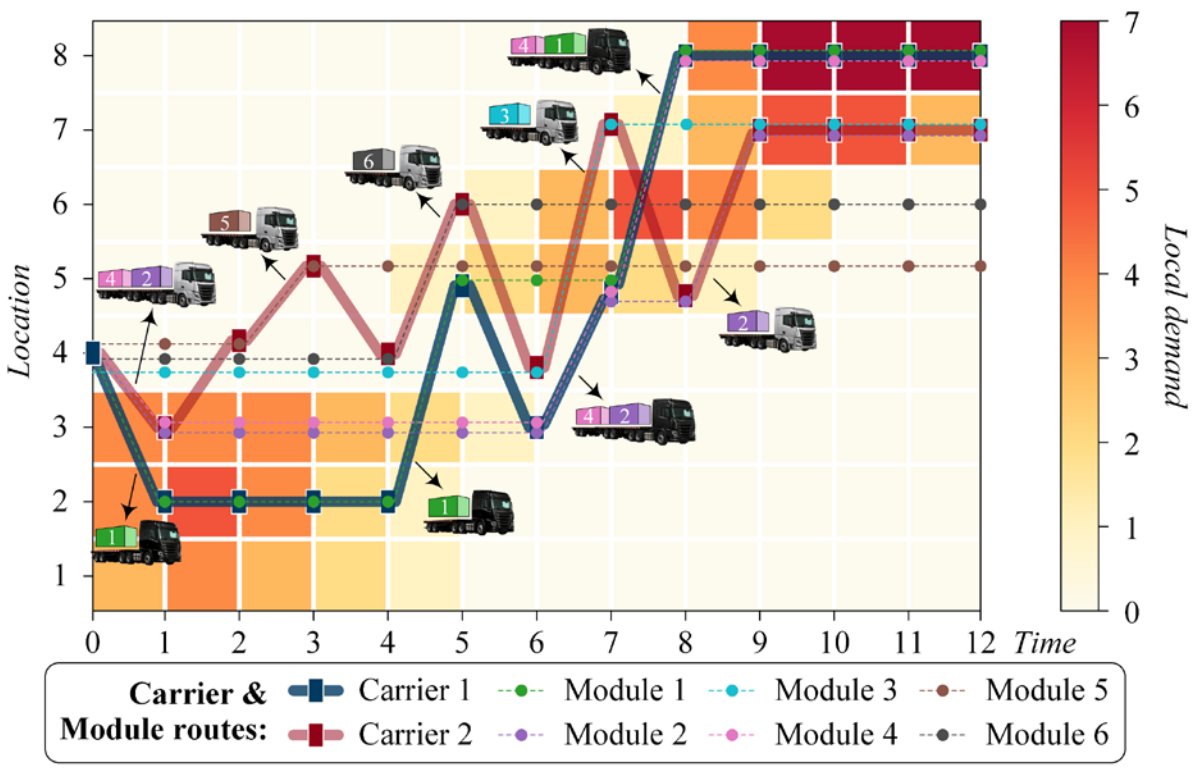

Fig. 3. Scheduling results of separable MERs.

To ensure a fair comparison with the logic-based benchmark in [9], we test a dwelling-constrained TSN variant by imposing the same post-arrival dwelling constraint as in (12).

$$\sum_{a \in \delta_j^{\mathrm{T},-}(t,i)} y_{j,a} \le y_{j,((t,i),(t+1,i))} \ , \forall j \in \mathcal{J}, i \in \mathcal{N}_j^{\mathrm{C}}, t \in \mathcal{T} \setminus \{0, D\} \tag{12}$$

The comparison results are summarized in Table III. The proposed bi-layer TSN formulation (Bi-TSN) achieves much lower solution times and better scalability than the logic-based model in [9]. Although both formulations solve all small 1-carrier/3-module instances, the logic-based model quickly reaches a computational bottleneck as the number of carriers and modules increases. In contrast, the bi-layer TSN remains substantially more tractable by exploiting the network-flow structure of carrier-supported module routing. This improved tractability makes the bi-layer TSN a more scalable tool for practical MER scheduling, where multiple separable resources must be coordinated under tight emergency-response windows.

TABLE III. Computational Comparison with the Logic-Based Model

| $\lvert\mathcal{J}\rvert$/ $\lvert\mathcal{K}\rvert$ | **Model** | **$\lvert\mathcal{N}\rvert$=4 ($\mathcal{N}_j^{\mathrm{C}}$=$\mathcal{N}_k^{\mathrm{M}}$=$\mathcal{N}$)** | **$\lvert\mathcal{N}\rvert$=8** | **$\lvert\mathcal{N}\rvert$=16** | **$\lvert\mathcal{N}\rvert$=32** | **$\lvert\mathcal{N}\rvert$=64** | **$\lvert\mathcal{N}\rvert$=128** |
|---|---|---|---|---|---|---|---|
| **1/3** | Bi-TSN | 0.16 | 0.30 | 1.16 | 2.39 | 6.53 | 27.13 |
| | Logic | 0.23 | 0.69 | 2.40 | 5.70 | 11.59 | 55.39 |
| **2/6** | Bi-TSN | 0.34 | 1.23 | 3.28 | 19.63 | 61.04 | 114.07 |
| | Logic | 5.20 | 33.05 | 177.01 | 370.38 | 600.22 | (**9/10**, 447.50, 0.76%) |
| **3/9** | Bi-TSN | 0.76 | 6.64 | 9.95 | 115.98 | 315.58 | 1155.50 |
| | Logic | 661.55 | (**3/10**, 1349.15, 0.76%) | (**2/10**, 1361.02, 2.25%) | (**3/10,** 1491.22, 3.52%) | (**0/10**, -, 3.89%) | (**0/10**, -, 4.10%) |
| **4/ 12** | Bi-TSN | 3.46 | 5.94 | 24.30 | 370.04 | (**7/10**, 491.71, 1.22%) | (**5/10**, 2020.44, 1.15%) |
| | Logic | (**2/10**, 987.58, 0.91%) | (**0/10**, -, 1.69%) | (**0/10**, -, 3.57%) | (**0/10**, -, 3.69%) | (**0/10**, -, 5.75%) | (**0/10**, -, 7.18%) |

**Note:** For each $\lvert\mathcal{N}\rvert$, a set of 10 demand instances is randomly generated and used for both models. Fully solved entries report average solution time in seconds; time-limited entries are given as "(solved instances, average solution time of solved instances, average final MIP gap of unsolved instances)".

## IV. Conclusion

Separable scheduling opens a flexible way to deploy MERs, yet an exact and tractable formulation is still needed to support its broader application. This letter addresses this need by recasting separable MER scheduling as a carrier-supported module routing problem and developing a bi-layer TSN formulation. By modeling carrier and module trajectories as interacting network flows with arc-support coupling, the formulation preserves exactness and eliminates mandatory post-arrival dwelling. Numerical results verify its correctness and demonstrate substantially improved solution efficiency, helping translate separable scheduling from a flexible concept into a practically usable optimization model.